\documentclass[fleqn,twoside]{article}
\usepackage{espcrc2}

\usepackage{graphicx}
\usepackage[figuresright]{rotating}

\newcommand{\AmS}{{\protect\the\textfont2
  A\kern-.1667em\lower.5ex\hbox{M}\kern-.125emS}}

\title{The relations between Bjorken polarized, Bjorken unpolarized,
and Gross--Llewellyn Smith sum rules}

\author{A.~L.~Kataev\address[MCSD]{Institute for Nuclear Research of 
the Russian Academy of Sciences,\\
117312 Moscow, Russia}
\thanks{Supported by RFBR Grants N 03-02-17177, 05-01-00992}}
\begin{document}

\begin{abstract}
New relations between Bjorken polarized, Bjorken unpolarized 
and Gross-Llewellyn Smith sum rules are described. These relations 
are valid in the  region, where both perturbative series 
and the series in power-suppressed $1/Q^2$-terms do  not yet 
manifest the feauture of asymtotic expansions. The experimentally based
consideartions support these relations, which may surve as the guide 
for possible in future measurements of the Bjorken unpolarized sum rule 
at Neutrino Factories. 
\vspace{1pc}
\end{abstract}
\maketitle

The dominating theoretical 
research  in the current studies  of  Neutrino 
Factories programs are  
related to the analysis of the possibility to detect parameters of 
neutrino oscilations. Since high-statistics measurments of 
cross-sections of $\nu$N-scattering 
with more intensive low-energu $\nu$- beams 
allow one to get more reliable estimates of background effects to 
future oscillations experiments, theoretical analysis of these processes 
started to attrackt more interest. 
Moreover, these studies 
may allow one to get new  information  
about 
the behaviour of cross-sections and extracted formfactors 
and  structure functions (SFs) 
in elastic, quasi-elastic and deep-inelastic scattering (DIS) 
regimes. 
In view of this even at low energies  the option 
for front-end non-oscillation physics should be  investigated in more 
detail. 
In particular, 
the analysis of the data in DIS region   may provide  high-presision 
information 
about polarized parton distributions \cite{Forte:2001ph},\cite{Mangano:2001mj} 
in the $x,Q^2$-regions, complementary to the ones, available 
at  JLAB.
Thus, using these parton 
distributions and  extrapolating the poissible Neutrino Factories 
data for $g_1(x,Q^2)$ SF  of polarized DIS one 
may extract the value for the isospin  polarized Bjorken sum rule 
\begin{equation} 
\rm{Bp}= \int_0^1\bigg[g_1^{p}(x,Q^2)-g_1^{n}(x,Q^2)\bigg]dx
\nonumber
\end{equation}
for the low  momentum transfered of about 
$Q^2\leq 3.5~{\rm GeV^2}$. In general QCD expression 
for the Bjp sum rule can be expressed as 
$\rm{Bp}=\frac{g_A}{6}\rm{C}_{\rm{Bp}})$
with $g_A$=1.26 being  the neutron beta-decay coupling constant 
and 
\begin{equation}
\rm{C}_{\rm{Bp}}=1-4a_s-O(a_s^2) - \frac{\rm{A}}{Q^2}- 
O\bigg(\frac{1}{Q^4}\bigg)
\end{equation}
where  $a_s$=$\alpha_s(Q^2)/(4\pi)$, $\alpha_s$ is the QCD coupling constant
and 
$\rm{A}$ is related to the non-perturbative $1/Q^2$-correction, 
calculated  numerically 
in different models (for the details see Ref. \cite{Kataev:2005hv}). 
Note, that  $1/Q^2$-corrections to $g_1^N$ were also extracted by 
the model-independent way  from the current 
data for polarized DIS \cite{Leader:2002ni}).
On another hand the kinenatical conditions of Neutrino Factories may allow one
to extract all SFs, which 
enter into the cross-section 
of the  unpolarized $\nu$N DIS process (i.e. $\rm{F}_1$, $\rm{F}_2$, 
$\rm{xF}_3$ SFs).
In view of this it may be possible to use data of Neutrino Factories 
for the {\bf first} extraction of unpolarized Bjorken sum rule, defined 
as
\begin{equation}
\rm{Bup}=\int_0^1\bigg[F_1^{\nu p}(x,Q^2)-F_1^{\nu n}(x,Q^2)\bigg]dx
\nonumber
\end{equation}
The QCD expression of $\rm{Bup}=\rm{C}_{\rm{Bup}}$, namely 
\begin{equation}
\rm{C}_{\rm{Bup}}=1-\frac{8}{3}a_s - O(a_s^2)- \frac{\rm{B}}{Q^2} - 
O\bigg(\frac{1}{Q^4}\bigg) ~~
\label{Bju}
\end{equation}
is ``measuring'' the {\bf violation} of the 
Callan-Gross relation from its parton model prediction $\rm{F_2/2xF_1}=1$.
The possibility to  test experimentally this property of QCD
through the extraction of Eq.(\ref{Bju})  
at Neutrino Factories was  proposed in Ref. \cite{Alekhin:2002pj}
(see also  \cite{Mangano:2001mj}).  

The third sum rule we are interested in is 
the Gross-Llewellyn Smith sum rule
\begin{equation}
\rm{GLS}=\frac{1}{2}\int_O^1\bigg[F_3^{\nu p}(x,Q^2)
+F_3^{\nu n}(x,Q^2)\bigg]dx~~. 
\nonumber
\end{equation}
It is proportional to the number 
of valence quarks, contained in the nucleon, namely 
$\rm{GLS}=3\rm{C}_{\rm{GLS}}(Q^2)$, where 
\begin{equation}
\rm{C}_{\rm GLS}=1-4s_s - O(a_s^2) -  \frac{C}{Q^2}- 
O\bigg(\frac{1}{Q^4}\bigg)~~.
\end{equation}
The key observation, which was made in Ref. \cite{Kataev:2005ci}, 
is that due to the fact that the first  infrared renormalon poles  
are entering into the 
Borel integrals for these sum rules 
with the same residues \cite{Broadhurst:2002bi}
\footnote{For the most recent review of the 
application of renormalon technique to  DIS 
sum rules see Ref. \cite{Kataev:2005hv}.}
not only  properly normalized perturbative contributions  
to all 
three sum rules have similar value
 \cite{Broadhurst:2002bi}, 
but the non-perturbative $1/Q^2$ corrections as well. Thus, in the energy 
region where asymptotic nature of the $1/Q^2$-expansion  did not yet  start to 
manifest itself (namely in the region $Q^2\geq 2 ~{\rm GeV^2}$), the following 
{\bf new relation }  between DIS sum rules 
\begin{equation}
\nonumber 
\rm {Bp}(Q^2)\approx \frac{g_A}{6}\rm{Bup}(Q^2) \approx \frac{g_A}{18}
\rm{GLS}(Q^2) 
\label{new}
\end{equation}
is  valid \cite{Kataev:2005ci}.
 It was checked that the existing 
experimental data for the GLS sum rule \cite{Kim:1998ki} 
and the existing data for the  Bp sum rule, extracted 
 both by experimentalists 
\cite{Abe:1998wq} and theoreticians 
\cite{Ellis:1995jv}, \cite{Altarelli:1996nm} 
are respecting the new relation of Eq.(\ref{new}) 
(see Ref. \cite{Kataev:2005ci}).
Indeed, using this relation one gets from the 
experimentally-based value $\rm{GLS}(Q^2=3.16~{\rm GeV^2})\approx 2.55$ 
is transformed into the value $\rm{Bp}(Q^2=3.15~{\rm GeV^2})\approx 0.178$.
Within existing error bars this result agrees with the 
value  $\rm{B}p(Q^2=3~{\rm GeV^2}) =  0.164\pm 0.023$, 
obtained by experimentalists  \cite{Abe:1998wq},
and with theoretically improved extractions 
and $\rm{Bp}(Q^2=3~{\rm GeV^2}) =  0.164\pm 0.011$ \cite{Ellis:1995jv},
$\rm{Bp}(Q^2=3~{\rm GeV^2}) =  0.177\pm 0.018$  \cite{Altarelli:1996nm}.  
Other examples of the validity of the 
relations of Ref. \cite{Kataev:2005ci}  were presented 
in more detail work of Ref.  \cite{Kataev:2005hv}. 
It is also interesting to note, that the analysis of 
of existing experimental 
data,  performed in Ref. \cite{Sidorov:2005mw},  
supports the relations between $1/Q^2$ corrections to $xF_3$ 
and $g_1^{p}-g_1^{n}$, which follow 
from the relations of Ref. \cite{Kataev:2005ci}, in the region  $x>0.2$, 
where the contributions 
of high-twist terms should be essentially important. 

To conclude, we hope  that the relation 
of Eq.(\ref{new})
will allow to test self-consistency  of 
extracting 
discussed   
sum rules values from  
the data of possible front-end DIS experiments 
at future Neutrino Factories.

{\bf Acknowledgements} 
This talk was prepared during the visit to ICTP (Trieste). 
I am grateful to the staff of this Center for 
providing excellent conditions for work. It is the pleasure to thank 
organizers of NuFact05 Workshop (Frascati) 
and V. Palladino in particular, 
for invitation and hospitality
at this interesting Workshop. I am also grateful to S.I. Alekhin 
and  D. J.  Broadhurst 
for productive collaboration.

\end{document}